# AB-Stacked Multilayer Graphene Synthesized *via* Chemical Vapor Deposition: A Characterization by Hot Carrier Transport

Carlos Diaz-Pinto [a,b,†], Debtanu De [a,b,†], Viktor G. Hadjiev [b], Haibing Peng [a,b,*]

[a] Department of Physics and [b] the Texas Center for Superconductivity, University of Houston, Houston, Texas 77204

[†] Authors contributed equally to this work.

[*] Corresponding author: haibingpeng@uh.edu

ABSTRACT

We report the synthesis of AB-stacked multilayer graphene *via* ambient pressure chemical vapor deposition on Cu foils, and demonstrate a method to construct suspended multilayer graphene devices. In four-terminal geometry, such devices were characterized by hot carrier transport at temperatures down to 240 mK and in magnetic fields up to 14 T. The differential conductance ($dI/dV$) shows a characteristic dip at longitudinal voltage bias $V$=0 at low temperatures, indicating the presence of hot electron effect due to a weak electron-phonon coupling. Under magnetic fields, the magnitude of the $dI/dV$ dip diminishes through the enhanced intra-Landau level cyclotron phonon scattering. Our results provide new perspectives in obtaining and understanding AB-stacked multilayer graphene, important for future graphene-based applications.

The synthesis of graphene on copper foils by chemical vapor deposition (CVD) has attracted considerable attention because of its potential in practical applications. [1-6] The quality of CVD graphene[3] has yet to match that of the graphene mechanically exfoliated from pristine graphite, [7,8] and therefore the CVD growth techniques are still under very active research. Various thermodynamic conditions involving both low pressure and ambient pressure are being addressed. In Ref. 9, the role of diffusion kinetics was discussed, and it was pointed out that at low pressure the synthesis process is self-limited with respect to the methane (which serves as the carbon source) flow rate, as opposed to the ambient pressure case. At ambient pressure, the mass transport of the carbon species to the catalyst surface dominates the synthesis process and produces multilayer graphene. [9] However, for multilayer graphene, the CVD method usually leads to a turbostratic stacking order [10] (*i.e.*, the layers are randomly oriented with respect to one another). To our knowledge, there has been no report on the CVD synthesis of continuous multilayer graphene films with AB-stacking order. [11] Recently, bilayer graphene with AB-stacking has been achieved by CVD with low hydrogen flow rate or no hydrogen, [12,13] and reduced hydrogen concentration has also led [14] to the improvement of the quality of CVD graphene. It has been further realized [15] that the absence of hydrogen can change the chemical kinetics of the decomposition of methane into active carbon and hydrogen. Therefore, the hydrogen concentration may play a role in determining the active carbon species at different stages of growth [15] and hence may affect the growth rate, the quality, and the stacking order of multilayer graphene.

Here, the synthesis of AB-stacked multilayer graphene on Cu foils through ambient pressure CVD is demonstrated. We have achieved continuous multilayer graphene films with AB stacking by using a high methane flow rate and eliminating the hydrogen during the growth. This finding is important not only by shedding light on the CVD synthesis mechanism, but also for offering controlled production of multilayer graphene with AB-stacking by CVD. The multilayer graphene is further used to fabricate suspended multilayer graphene devices suitable for transport measurements. Hot electron transport is investigated through a four-terminal configuration as a function of the measured longitudinal voltage drop $V$, temperature $T$ and magnetic field $B$. Such four-terminal transport results unambiguously demonstrate a distinct differential conductance dip at $V = 0$, an intrinsic transport behavior of multilayer graphene attributable to the presence of hot electron effect. We further observed that magnetic fields attenuate the hot electron effect and thus provide a way for its control by enhancing cyclotron-phonon scattering. [16, 17]

## Results and Discussion

The growth of multilayer graphene was carried out on 25 μm thick Cu foils, using an ambient pressure methane CVD technique (see the Methods section for details of the CVD process). In brief, the graphene synthesis is realized in a key step with a mixture of a methane flow (13 mL/min) and an argon flow (150 mL/min) at a temperature of 1000 °C in a time period ~10 minutes, while warming up is done under a pure hydrogen flow (200 mL/min) and cooling down is done under a pure Ar flow (150 mL/min). We find that increasing the methane flow (> 10 mL/min) during the growth results in thicker multilayer graphene films, while a reduced methane flow (~1 mL/min) leads to single layer graphene. In this study, we focus on devices made of multilayer graphene films ~ 3 nm (or ~ 9 layers) in thickness.

After CVD, thin flakes of multilayer graphene are obtained by etching away the underlying copper substrates with iron nitride solution. They are then transferred to a 200 nm $SiO_2$/Si substrate for Raman characterization. Figure 1a shows the Raman spectra (514 nm $Ar^+$ laser line) for a multilayer graphene sample. The most important feature observed in the Raman spectra is the asymmetry of the $G'$ band at ~2730 $cm^{-1}$ (with more weight in the higher-frequency part), which is the signature of AB-stacking in multilayer graphene.[18-21] In contrast, for a multilayer graphene with a turbostratic stacking order, as typically reported in previous CVD work,[10] the Raman spectra should have a symmetric $G'$ band. The $G'$ band can be fitted by three Lorentzian curves (Fig. 1a, line) according to the methodology of Ref. 19. From the Raman intensity of the decomposed $G'_2$ and $G'_3$ bands (Fig. 1a), $I_{G'2}$ and $I_{G'3}$, we find a $c$-axis crystalline coherence length $L_c$ = 10 + 10/(1.05-$A$) ≈ 29 nm, where $A$ = $I_{G'3}$/($I_{G'2}$ + $I_{G'3}$). The estimated coherence length $L_c$ is longer than the thickness of the multilayer graphene (~3 nm measured by an atomic force microscope), indicating single-crystal quality along the $c$-axis and supporting the observation of AB-stacking from Raman measurements. The Raman spectra also reveals the presence of the disorder induced bands $D$ and $D'$ at ~1364 $cm^{-1}$ and ~1626 $cm^{-1}$, respectively (Fig. 1a inset). From the relative intensity of the $G$ band at ~1586 $cm^{-1}$ to the $D$ band, we can estimate the in plane crystallite size [18] $L_a$ = 16.6 ($I_G$/$I_D$) ~ 45 nm for this sample. We stress that in the CVD process, the elimination of hydrogen gas flow during the growth stage is critical for the synthesis of the AB-stacked multilayer graphene. To demonstrate the effect of the forming gas during growth, we performed controlled experiments for two CVD growth processes with and without a hydrogen gas flow at 50 mL/min in addition to a methane flow at 13 mL/min during the growth stage and compared their Raman spectra (Fig. S1 of the Supporting Information), which clearly demonstrates that the hydrogen gas flow leads to a symmetric $G'$ peak (indicating turbostratic stacking) and a larger defect-induced $D$ peak.

By using the etching and the transfer process, we are able to produce continuous multilayer graphene films in a size of a few millimeters, which is convenient for making devices for next-stage characterizations. Fig. 1b presents an optical image of such a multilayer graphene film placed on a $SiO_2$/Si substrate. We have also performed Raman mapping experiments (Fig. S2 of the Supporting Information) to further exam the uniformity of the films, which shows that the typical graphene film after etching and transferring to $SiO_2$/Si substrate is uniform in a scale of a few μm though shows thickness variation in larger scale.

In order to characterize the electronic properties of such CVD synthesized AB-stacking multilayer graphene, we further construct devices with a multilayer graphene flake suspended on top of six metal electrodes (see Methods section for device fabrication details). A schematic of the designed devices is shown in Fig. 1c, along with a scanning electron microscope (SEM) image of a device partially covered by a suspended graphene flake (Fig. 1d). The gap separating adjacent electrodes is ~400 nm wide, and the width of electrodes is ~1 μm. The available six electrodes allow multiple possible arrangements to be used in a four terminal measurement, but in experiments we select those four electrodes that have the lowest contact resistance to the graphene. The differential conductance $dI/dV$ vs. longitudinal voltage drop $V$ is measured *via* a standard lock-in technique by supplying a DC current with a small AC modulation (503 Hz) through the outer I+ and I- electrodes and measuring both the DC and AC voltage drop between two inner V+ and V- electrodes. The *dI/dV* is also measured as a function of the gate voltage $V_g$ with the degenerately doped silicon serving as a back gate (Fig. 1c).

The *dI/dV* as a function of gate voltage $V_g$ (Fig. 2a) shows a maximum tuning of ~50% with respect to the minimum conductance point within a $V_g$ range of ±50V at T = 240 mK, for a section of the multilayer graphene of Fig. 1d.[22] The minimum conductance point (neutral point) of the multilayer graphene is reached at a positive gate voltage $V_g$ = 24 V, indicating a typical hole-doped (*p*-type) behavior which can be induced by the adsorption on the graphene surface.[23]

The dependence of the *dI/dV* on the measured DC voltage *V* at T = 240 mK is shown in Fig. 2b for fixed $V_g$ = 0. The intriguing feature is a dip of *dI/dV* at *V* = 0 with a magnitude of ~ 40% of the zero-bias *dI/dV* value. Previously, we also observed such a dip in suspended graphitic multilayers by two-terminal transport measurements[16] for which the role of the contact resistance had to be carefully considered. Here our four-terminal transport results unambiguously demonstrate that the observed *dI/dV* dip is an intrinsic transport behavior of multilayer graphene with no contribution from the contact barrier. As seen in our experiments (Fig. S3 of the Supporting Information), the *dI/dV* dip remains at bias *V*=0 regardless of the value of the gate voltage, thus ruling out its relation to graphene band features (*e.g.* a neutral point), Coulomb blockade effect in an open quantum dot, Fabry-Perot resonance, or a possible presence of electron-hole puddles that partly form tunnel junctions.

Figure 3a presents the *dI/dV* vs. *V* with $V_g$ = 0 at various temperatures from 240 mK to 50 K (from bottom to top offset vertically for clarity). As the temperature is increased, the magnitude of the *dI/dV* dip is gradually suppressed, and completely disappears for temperatures above *T*=8 K. Figure 3b shows the *dI/dV* plotted as a function of temperature *T* for fixed *V* values. The temperature dependence of the *dI/dV* presents a similar behavior as the *dI/dV* vs. *V* curve, *i.e.*, a dip of *dI/dV* is observed for *V* = 0 at low temperatures. This dip feature vanishes gradually as the bias *V* increases. The similar behavior of the *dI/dV* as a function of *T* and *V* suggests the presence of electron heating, and the *dI/dV* dip can be related to the hot electron effect [16, 24, 25] as explained below. At finite bias, the energy supplied to the charge carriers by the electric field cannot be transferred to the lattice fast enough due to a relatively weak electron-phonon scattering at low temperatures, resulting in an effective electron temperature $T_e$ higher than the lattice temperature $T_l$ (same as the cryostat temperature *T* in experiments). Therefore, the *dI/dV* value for a finite bias *V* measured at the cryostat temperature *T* should be equal to the *dI/dV* value for zero bias at an effective temperature $T_e$ higher than *T*. Consequently, the experimentally obtained *dI/dV* vs *V* at a fixed low temperature (*T* < 8 K in Fig. 3) should follow the trend of *dI/dV* vs *T* at zero bias (*V*=0). This explains why the magnitude of the dip at zero bias in *dI/dV* vs. *V* is suppressed as the temperature is increased (Fig. 3a) [16]. At high enough temperatures (*T* > 8K for the case of Fig. 3), the achievable effective hot carrier temperature $T_e$ is no longer higher than that of the lattice temperature due to enhanced electron-phonon scattering at high *T*, and therefore the *dI/dV* dip disappears (Fig. 3a). [26]

In magnetic fields, the *dI/dV* as a function of *V* presents two notable changes (Fig. 4). First, the overall *dI/dV* values drop with increasing magnetic fields (Fig. 4a), indicating the presence of a strong positive magnetoresistance (MR) at high magnetic fields, which is commonly observed in bulk graphite [27, 28] or graphitic multilayers [29] and can be attributed to the interplay between long-range and short-range disorders. [29] At low magnetic fields (Fig. 4b), negative magnetoresistance appears at low temperatures due to the weak localization effect, which indicates the influence of defects including grain boundaries in CVD-grown graphene. [30] Second, as the magnetic field is increased, there is an attenuation of the hot electron dip of *dI/dV* at *V*=0 (Fig. 4a), and this phenomenon is reproducibly observed in different samples (see Fig. S4 of the Supporting Information). The magnitude of the *dI/dV* dip is considerably reduced at 14 T, as compared with the case at zero field. However, the persistence of the *dI/dV* dip even at 14 T suggests that weak localization is not the dominant source for the origin of the *dI/dV* dip in this sample. Otherwise, the dip should vanish at lower *B* fields [16, 31]. The *B* field induced suppression of the *dI/dV* dip observed here is more likely related to the enhanced electron-phonon scattering at high *B* fields [17] which diminishes the hot electron effect and thus the *dI/dV* dip. Under high magnetic fields, the intra-Landau level electron-phonon scattering is enhanced as a result of the increased degeneracy for a Landau level (equal to $B/\varphi_0$ with $\varphi_0 = h/e$), making it easier for electrons to lose the energy gained from the electric field to the lattice. As a consequence, the effective electron temperature that can be reached at high magnetic fields is lower than that achievable at zero field, leading to the attenuation of the hot electron dip near *V* = 0 (Fig. 4a).

Moreover, we have observed similar gate tuning behavior, the persistence of the zero bias *dI/dV* dip and its attenuation under magnetic fields for different sections of the multilayer graphene of Fig. 1d (see Fig. S4 of the Supporting Information), indicating the homogeneity of the electron transport properties in the as-prepared multilayer graphene.

**Conclusion**

We have succeeded in the synthesis of AB-stacked multilayer graphene *via* a distinct atmospheric pressure CVD technique, as confirmed by Raman scattering experiments. Further investigation of its electronic properties is performed by the fabrication and electron transport studies of four-terminal multilayer graphene devices. The *dI/dV* as a function of longitudinal voltage bias *V* shows a dip pinned at *V* = 0, [32] indicating the presence of hot electron effect due to a weak electron-phonon coupling. Under magnetic fields, the magnitude of the *dI/dV* dip diminishes through the enhanced intra-Landau level cyclotron phonon scattering. Our results provide new perspectives in obtaining and understanding AB-stacked multilayer graphene, important for future graphene-based applications.

**Methods**

The CVD growth was carried out under ambient pressure in a 1" quartz tube furnace, using copper foil 25 μm thick (Alpha Aesar, item no. 13382). At room temperature, argon gas is used to flush the quartz tube at a rate of 2 L/min for at least 15 minutes. The Ar flow is then stopped, a hydrogen flow of 200 mL/min is started and the temperature is raised to 1000 °C in ~18 min. With a subsequent waiting of 20 min at 1000 °C, the $H_2$ flow is stopped, and an Ar flow of 150 mL/min is initialized and maintained for another 20 min. After that, a methane flow is established at 13 mL/min together with the Ar flow at 150 mL/min, followed by a waiting time of 10 min (typical target time for graphene growth). In the end, the CH$_4$ flow is turned off and the system is cooled down to room temperature in ~ 1.5 hrs while keeping the Ar flow at 150 mL/min.

The following steps are needed for the fabrication of suspended multilayer devices after the CVD growth on copper foils. First, one side of the graphene-covered Cu foils is protected by spin coating of PMMA (~200 nm thick), while the graphene film on the other side of Cu foils is etched away using reactive ion etching with 50 sccm of O$_2$ at 50 W for 2 min. The obtained PMMA/graphene/copper foil is then cut into small ribbons (~2x5 mm) and placed in a Fe(NO$_3$)$_3$ aqueous etching solution (0.2 g/mL) for a few hrs. After the copper is etched, the PMMA/graphene membrane is rinsed three times in deionized (DI) water, followed by a transfer to a SiO$_2$/Si substrate with a pre-patterned array of electrode sets (Fig. 1c). The sample is then blow dried with N$_2$ gas and kept in a vacuum desiccator overnight, and thus the membrane is strongly adhered to the substrate. Subsequently, the PMMA layer is removed by acetone at ~ 55 $^o$C for 30 min. Finally, ebeam lithography and reactive ion etching are employed again to define a rectangular shape of the graphene suspended over electrodes (Fig. 1c), followed by a lift-off process in Acetone to remove the PMMA residue from the last lithography step. Measurements by atomic force microscope on typical devices show that the multilayer graphene sheets are suspended as designed (Fig. S5 of the Supporting Information).

For the device of Fig. 1d, we performed electrical measurements on the as-prepared sample without thermal annealing. Low temperature measurements were carried out in a $^3$He fridge at a base temperature of 240 mK. Magnetic fields are applied perpendicular to the multilayer graphene with a superconducting magnet inside a cryostat. The differential conductance *dI/dV* vs. longitudinal voltage drop *V* is measured *via* a standard lock-in technique by supplying a DC current with a small AC modulation (503 Hz) through the outer I+ and I- electrodes and measuring both the DC and AC voltage drop between two inner V+ and V- electrodes. In the measurements for Figs. 2-4, we used one AC current source with an internal resistance of 1MΩ superimposed to a DC current source with an internal resistance of 20 KΩ to obtain the differential conductance *dI/dV* as a function of *V* or $V_g$. We note that part of the AC current goes through the internal resistance of the DC current source so that the actual AC current *dI* through the graphene is smaller than the nominal value provided by the AC current source. For the device of Figs. 2-4, the load resistance $R_{load}$ is 1673 Ω including circuit and contact resistance in series with the target graphene resistance (250 Ω at $V_g$ =0). The maximum variation of the graphene resistance as a function of *V* or $V_g$ is no more than 100 Ω, much less than the internal resistance of both current sources. Thus the actual *dI/dV* can be obtained accurately by multiplying a scaling factor calculated based on the load resistance (our controlled experiments with a single current source to measure *dI/dV vs.* $V_g$ have confirmed this). All the *dI/dV* values shown in this work are the accurate values after the correction.

The Raman measurements were done using a Horiba, JY Raman microscope equipped with a tunable $Ar^+$ laser set to operate at 514 nm in order to compare the measured spectra directly with those reported in literature, unless the laser wavelength is specified for certain experiments.

**Supporting Information Available:** Additional experimental results are presented on the effect of hydrogen during the graphene growth, Raman mapping of a graphene film, the differential conductance *dI/dV* as a function of bias *V* with different fixed gate voltages $V_g$, electron transport data for different sections of the device under study in the main text, and the AFM image of a suspended multilayer graphene device. This material is available free of charge *via* the Internet at http://pubs.acs.org.

FIGURE CAPTIONS

**Figure 1.** (a) Main panel: Raman $G'$ band measured with a 514 nm laser (symbols) and the best fit to the data (solid line) for a CVD-grown multilayer graphene flake transferred onto a SiO$_2$/Si wafer. The curve of the best fit is decomposed into three Lorentzian peaks $G'_1$, $G'_2$, and $G'_3$ centered at 2688 cm$^{-1}$, 2706 cm$^{-1}$, and 2734 cm$^{-1}$, respectively. Inset: Measured $D$, $G$ and $D'$ bands. (b) Optical microscope image of a multilayer graphene sample transferred to a SiO$_2$/Si substrate with a 200 nm oxide layer. Scale bar: 1 mm. (c) Schematic cross section (top panel) and top view (bottom panel) of a multilayer graphene device (not to scale). The dark gray rectangle (bottom panel) on top of the predefined electrodes represents the multilayer graphene flake designed to be 44 μm in length and 30 μm in width. The metal electrodes are 60 nm in thickness (52.5 nm Pd on top of 7.5 nm Cr). The gap separating adjacent electrodes is ~ 400 nm wide. The width of the electrodes is ~ 1 μm for the four electrodes in the center and ~ 4μm for the two at the edges. (d) Scanning electron microscope image of the device used in Figs. 2-4. The multilayer graphene lying on top of the metal electrodes can be seen with a light contrast change (in the lower part of the image). Scale bar: 2 μm.

**Figure 2.** (a) Differential conductance $dI/dV$ as a function of the gate voltage $V_g$ at fixed longitudinal voltage bias $V = 0$. (b) The $dI/dV$ as a function of the measured longitudinal voltage bias $V$ with $V_g = 0$. All data are taken at a temperature $T = 240$ mK.

**Figure 3.** (a) Differential conductance $dI/dV$ as a function of the measured longitudinal voltage bias $V$ with $V_g = 0$ at different temperatures (from bottom to top: T = 240 mK, 2.5, 4.5, 6, 8, 10, 15, 36 and 50 K). For clarity, the curves are stacked by a vertical offset of 0.15 mS from bottom to top. (b) $dI/dV$ data plotted as a function of temperature $T$ for selected values of bias $V$. The curves are offset by 0.1 mS for clarity (from bottom to top: $V = 0$, -0.55, -3.0, and -8.0 mV).

**Figure 4.** (a) Differential conductance $dI/dV$ as a function of $V$ with $V_g = 0$ V in different magnetic fields at a temperature 240 mK. The curves are plotted without any vertical offset. (b) Magnetoresistance data with fixed longitudinal bias $V = 0$ and gate voltage $V_g = 0$ at different temperatures. The weak localization effect is attenuated by increasing temperatures.

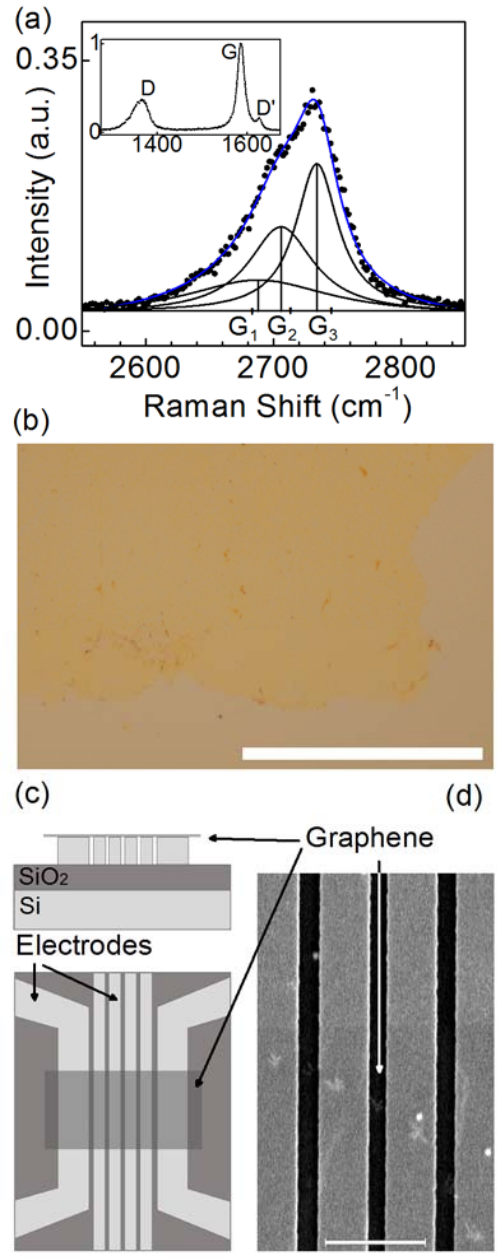

Fig. 1

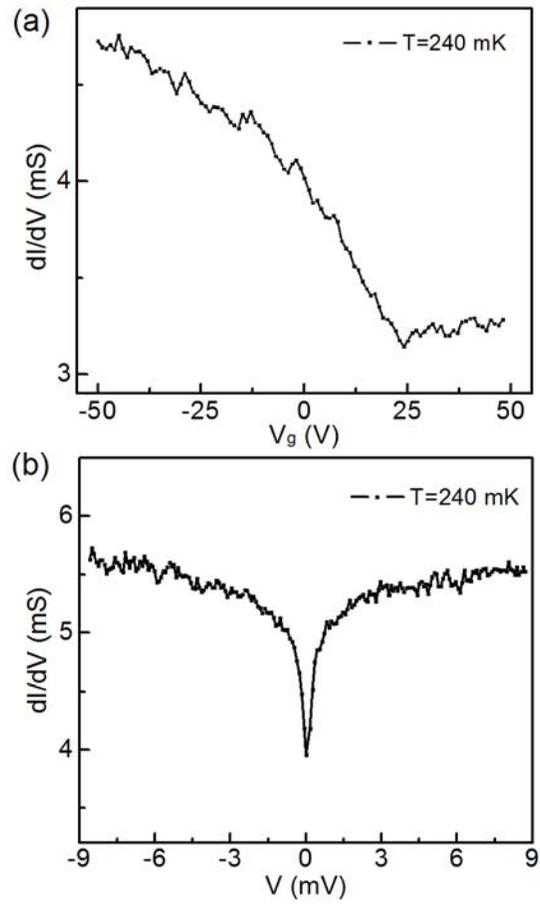

Fig. 2

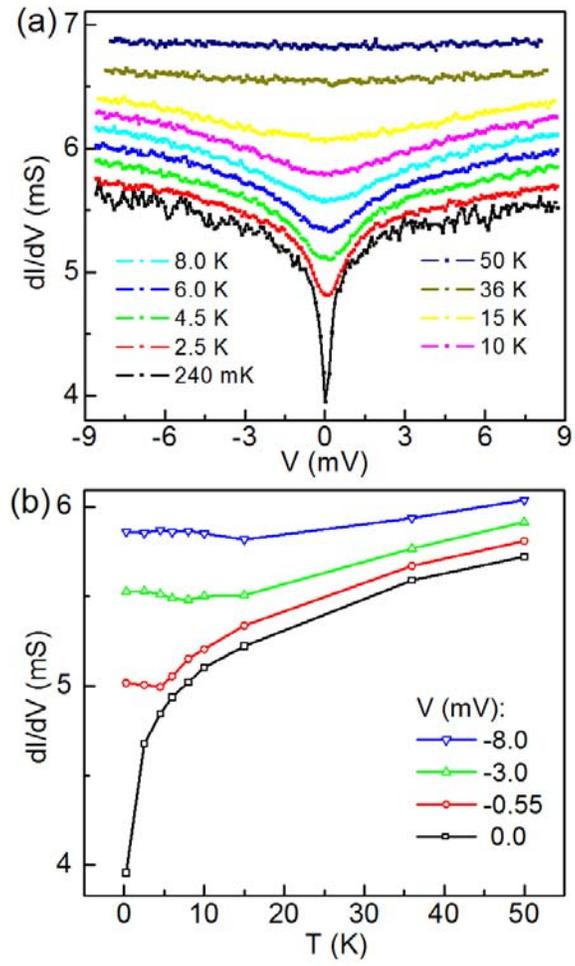

Fig. 3

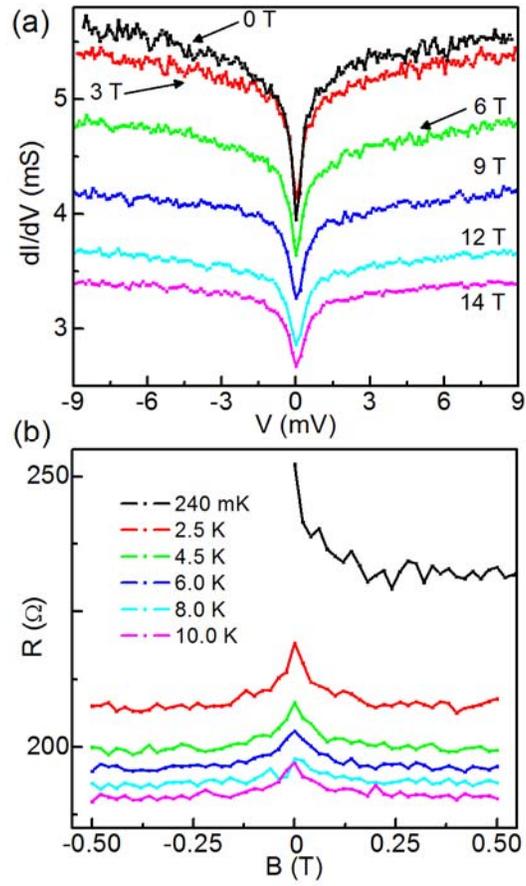

Fig. 4

# Supporting Information for:

# AB-stacked multilayer graphene synthesized via chemical vapor deposition: a characterization by hot carrier transport

Carlos Diaz-Pinto [a,b,†], Debtanu De [a,b,†], Viktor G. Hadjiev [b], Haibing Peng [a,b,*]

[a] Department of Physics and [b] the Texas Center for Superconductivity, University of Houston, Houston, Texas 77204

[†] Authors contributed equally to this work.

[*] Corresponding author: haibingpeng@uh.edu

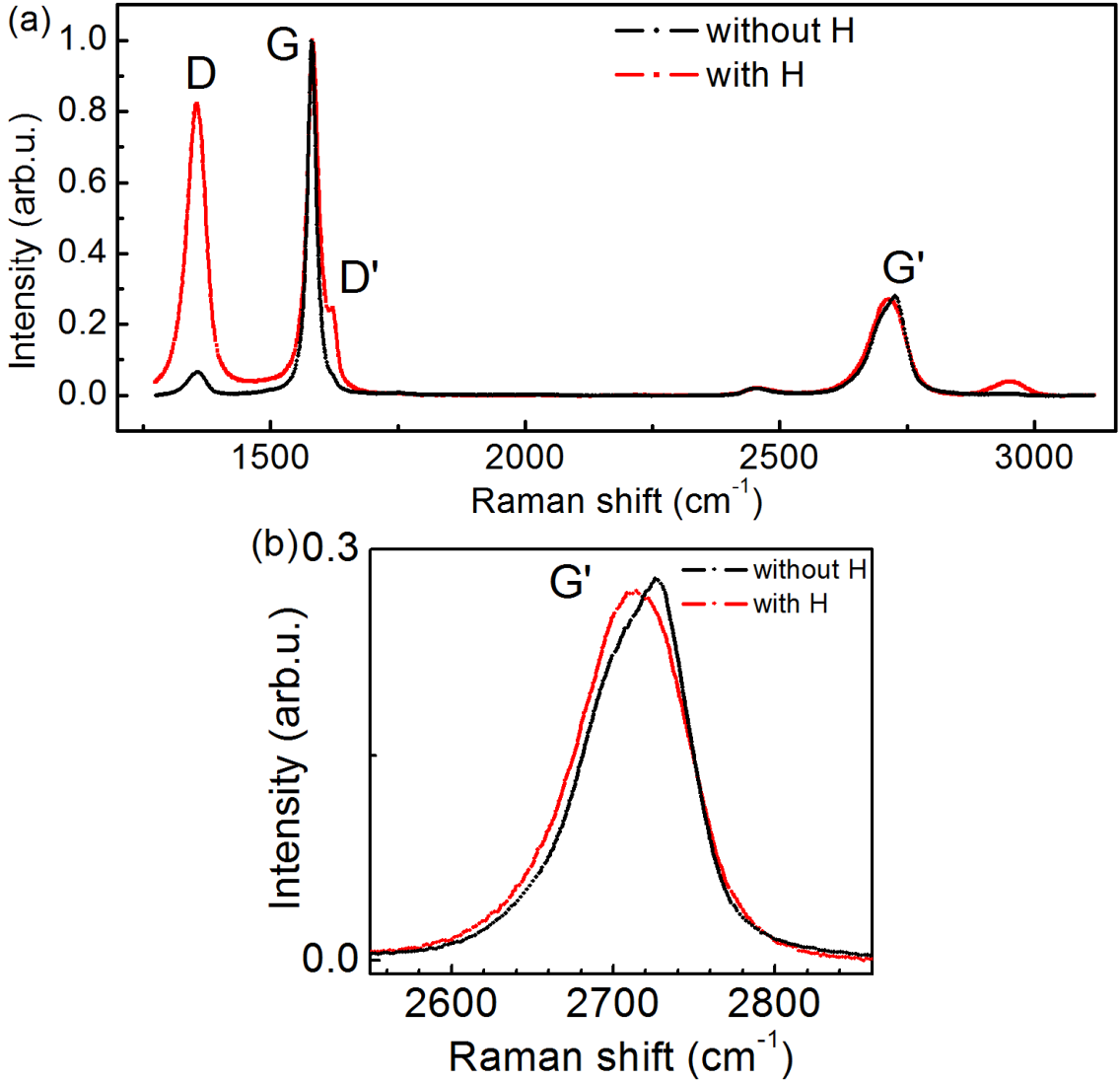

**Figure S1.** (a) Raman-scattering spectra of multilayer graphene samples grown without hydrogen flow (black curve) and with a hydrogen flow of 50 mL/min (red curve) during the growth stage. In the former case, the recipe is the same as described in the method section of the main text, i.e. a methane flow of 13 mL/min and an Ar flow of 150 mL/min were used during the growth stage, but with a growth time of 20 min. In the latter case, a methane flow of 13 mL/min, an Ar flow of 100 mL/min, and a hydrogen flow of 50 mL/min were used during the growth stage with a growth time of 20 min. The multilayer graphene was transferred to a $SiO_2$/Si substrate with a 200 nm oxide layer prior to the Raman characterization. (b) A zoom-in view of the $G'$ peaks in (a).

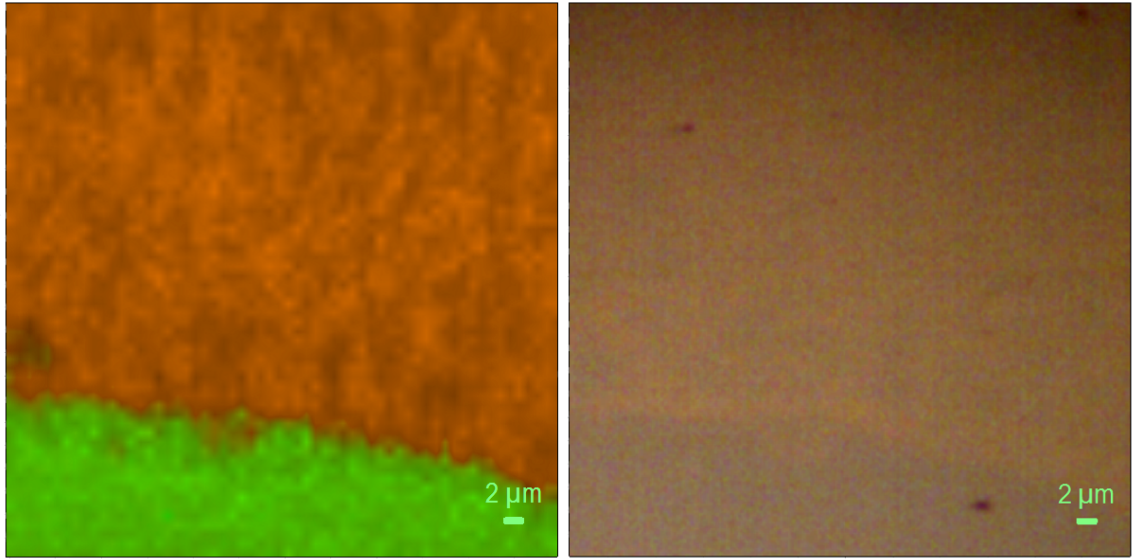

**Figure S2.** Raman mapping image (left) and the corresponding optical image (right) of a multilayer graphene sample (transferred to a 200 nm $SiO_2$/Si substrate). A laser with the wavelength 632.8 nm was used for the experiments. The multilayer graphene film is uniform in a scale of a few μm though shows thickness variation in larger scale.

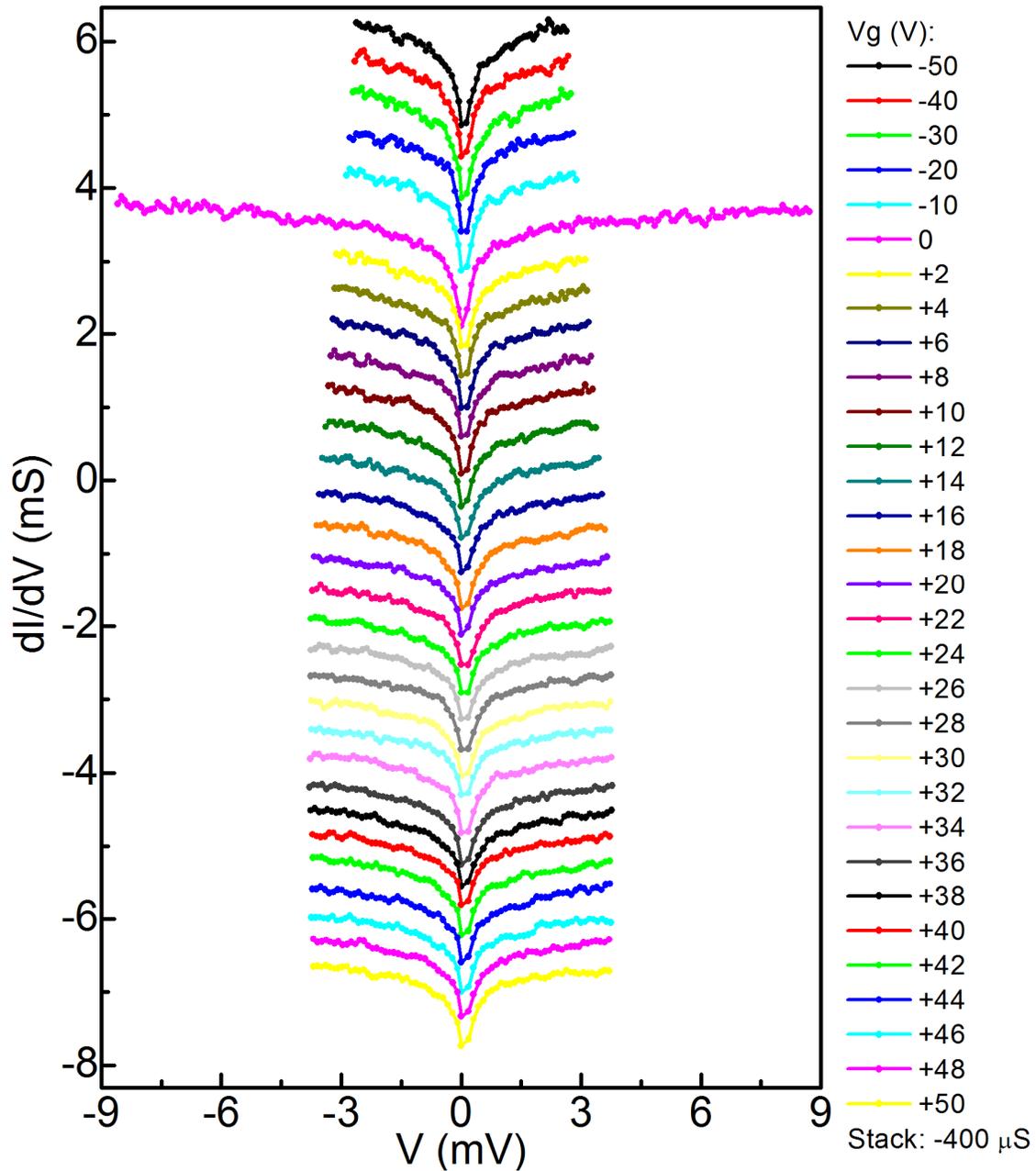

**Figure S3.** Differential conductance $dI/dV$ as a function of the measured longitudinal voltage $V$ with various gate voltage $V_g$ at $B = 0$ T and $T = 240$ mK. For clarity, the curves are stacked with an offset - 400 μS from the top curve ($V_g = -50$ V) to the bottom one ($V_g = +50$ V).

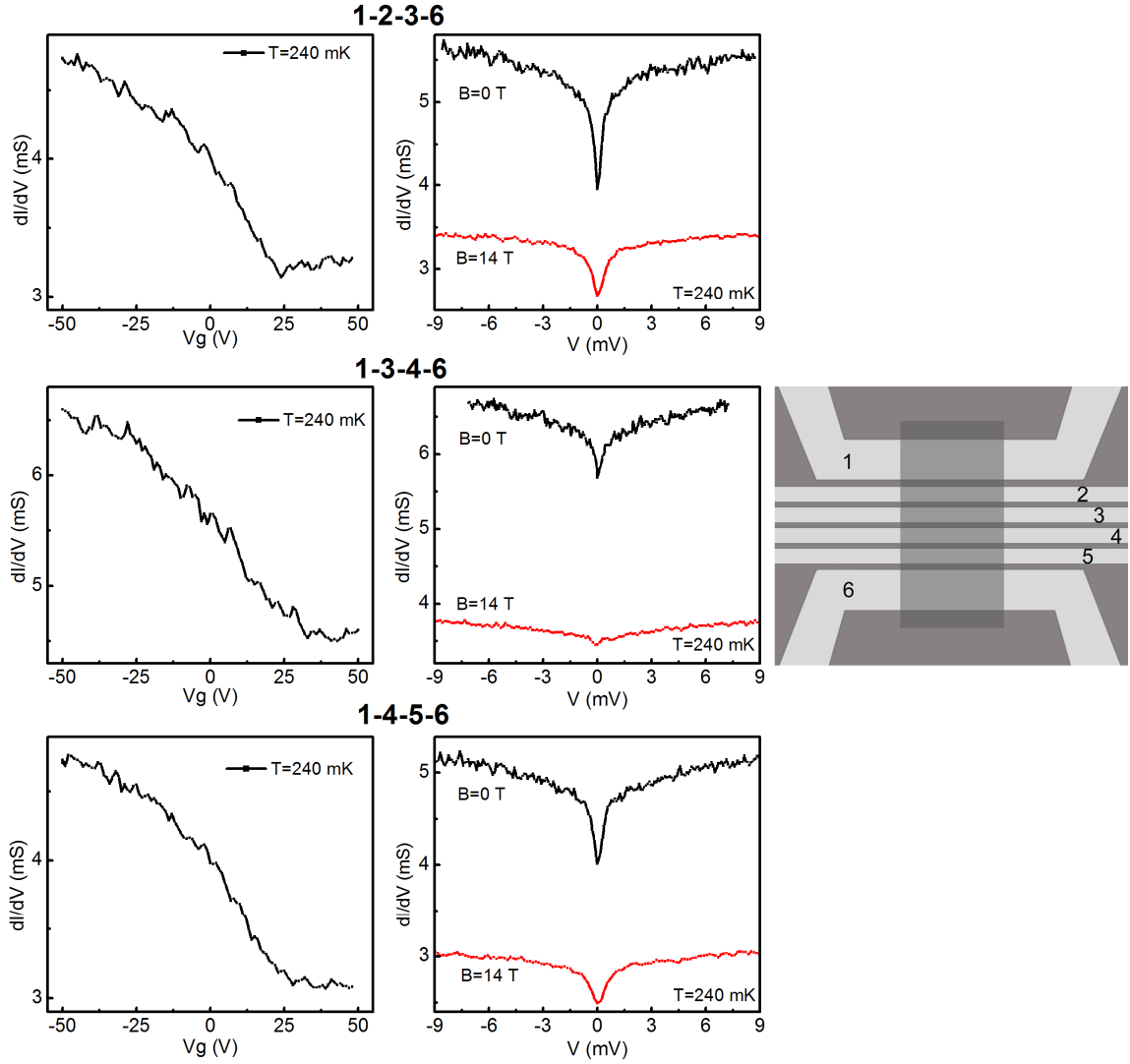

**Figure S4.** Differential conductance *dI/dV* as a function of gate voltage $V_g$ at bias $V = 0$ (left column) and *dI/dV* as a function of *V* at $V_g = 0$ in magnetic fields *B*=0 and *B*=14 T (center column) at *T* =240 mK for three different segments of the same multilayer graphene sample under study in the main text. The schematic and electrode labeling of the device is shown on the right column. In all three cases, the current is run from *I+* (electrode #1) to *I-* (electrode #6), while both the AC and DC voltage drops are measured between *V+* and *V-* (electrodes #2 and #3 for the data in row 1, #3 and #4 for the data in row 2, and #5 and #6 for the data in row #3). Different configurations of electrodes in the measurements are labeled by the electrode numbers in a sequence of $I_+$ - $V_+$ - $V_-$ - $I_-$. The configuration 1-2-3-6 (the top row) is the one used for Figs. 2-4 of the main text.

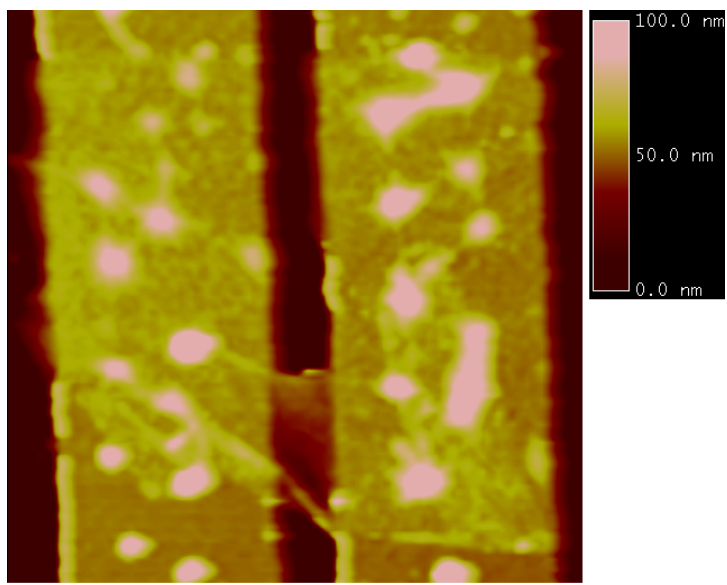

**Figure S5.** Atomic force microscope (AFM) image of a typical device with suspended multilayer graphene bridging two electrodes (we note that suspended multilayer graphene device can be easily perturbed by an AFM tip during the imaging process). The size of the image is 2.77 μm x 2.77 μm. The maximum height in the scale is 100 nm.